\documentclass[doublecol]{epl2}
\title{The predictions of the charm structure function exponents behaviour at low $x$ in deep inelastic scattering }
\author{G.R.Boroun\inst{1} \and B.Rezaei\inst{1}}

\institute{
  \inst{1} Physics Department, Razi University, Kermanshah
67149, Iran} 

\abstract{ We use the hard (Lipatov) pomeron for the low
 $x$ gluon distribution and provide a compact formula for the
 ratio
 $R^{c}=\frac{F^{^{c\overline{c}}}_L}{F^{^{c\overline{c}}}_2}$
 that is useful to extract the
 charm structure function from the reduced charm cross section, in particular at DESY HERA.
 Our results show that this ratio is  independent of $x$ and independent of the DGLAP evolution of the gluon PDF.
As a result, we show that the charm structure function and the
reduced charm cross section exponents do not have the same
behaviour at very low $x$. This difference is independent of the
input gluon distribution functions and predicts the non- linear
effects and some evidence for shadowing and antishadowing at HERA
and RHIC.}

\begin{document}

\maketitle

\section{1. INTRODUCTION}
Precise measurements of the charm inclusive scattering cross
section at the $ep$ collider are important for the understanding
of charmed meson  production. In the one-photon exchange
approximation the neutral current, charmed meson production in
deeply inelastic $ep$ scattering is via this reaction
\begin{equation}
e^{-}+P{\rightarrow}e^{-}+c\overline{c}+X\nonumber.
\end{equation}
In the case of pure photon exchange, the totally inclusive cross
section of the deep-inelastic lepton-proton scattering (DIS) has
the form:
\begin{equation}
\frac{d^2\sigma}{dx dQ^2}={\frac{2\pi{\alpha}^2 Y_+}{Q^4
x}}.\sigma_r,
\end{equation}
where the reduced cross section is defined as
\begin{equation}
\sigma_r\equiv F_2 (x,Q^2)-{\frac{y^2}{Y_+}}.F_L (x,Q^2),
\end{equation}
and $Y_+ =1+(1-y)^2$. Here $Q^2$ is the squared four-momentum
transfer, $x$ denotes the Bjorken scaling variable, $y=Q^2/sx$ is
the inelasticity, with $s$ the ep center of mass energy squared,
and $\alpha$ is the fine structure constant. The structure
functions\quad $F_2$\quad and\quad $F_L$\quad are related to the
cross sections $\sigma_T$ and $\sigma_L$ for interaction of
transversely and longitudinally polarized virtual photons with
protons [1]. At small values of $x$, $F_{L}$ becomes
non-negligible and its contribution should be properly taken into
account when the $F_{2}$ is extracted from the measured cross
section. However, the contribution of the longitudinal structure
function $F_L$ to the cross section is sizeable only at large
values of the inelasticity y, in most of the kinematic range the
relation $\sigma_r\approx F_2$ holds to a very good approximation.
The same is true also for the contributions
$F^{^{c\overline{c}}}_2$ and $F^{^{c\overline{c}}}_L$ to $F_{2}$
 and $F_{L}$ due to the charm quarks.\\

In perturbative QCD (pQCD) calculations, the production of heavy
quarks at HERA proceeds dominantly via the direct boson-gluon
fusion (BGF) where the photon interacts with a gluon from the
proton by the exchange of a heavy quark pair [2]. Charm production
contribution to the DIS cross section data was found to be around
$30\%$ $F_{2}$ at HERA [3,4]. The deeply inelastic heavy-flavour
structure function contribution to the cross section is given by
\begin{eqnarray}
\sigma^{c\overline{c}}_r&=&\frac{Q^4 x}{2\pi{\alpha}^2
Y_+}\frac{d^2\sigma^{c\overline{c}}}{dxdy}\nonumber\\
&&=F^{^{c\overline{c}}}_2
(x,Q^2,m^{2})-{\frac{y^2}{Y_+}}F^{^{c\overline{c}}}_L
(x,Q^2,m^{2})\nonumber\\
 &&=F^{^{c\overline{c}}}_2
(x,Q^2,m^{2})(1-{\frac{y^2}{Y_+}}R^{c}).
\end{eqnarray}

A measurement of the longitudinal charm structure function at low
$x$ at HERA is important because the $F^{^{c\overline{c}}}_L$
contribution to the charm cross section can be sizeable. At small
values of $x$, $F^{^{c\overline{c}}}_L$ becomes non-negligible and
its contribution should be properly taken into account when the
$F^{^{c\overline{c}}}_2$ is extracted from the measured charm
cross section. This has been done, for example for H1 charm data,
by making an NLO QCD fit in the DGLAP formalism such that the
correction necessary for $F_{L}^{c\overline{c}}$ is calculated in
this formalism. Instead of this procedure, we propose to use the
expression 4 with the quantity $R^{c}$ determined in NLO
approximation, which has the advantage of being independent of the
gluon PDF. This simplifies the extraction of
$F^{^{c\overline{c}}}_2$ from measurements of
$\sigma^{c\overline{c}}_r$.\\

In this paper, the charm structure function is evaluated from the
charm reduced cross section (4) where the ratio of the charm
structure functions, $R^{c}$, is calculated in a more robust NLO
approximation. We propose to use the  hard (Lipatov)Pomeron
behaviour of the charm structure functions  and determine the
ratio of the charm structure functions
$R^{c}=\frac{F^{^{c\overline{c}}}_L}{F^{^{c\overline{c}}}_2}$ from
this behaviour in the limit of low $x{\rightarrow}0$. Assuming the
low $x$ asymptotic behaviour of the gluon PDF  to be of the type
$G(x,Q^{2}){\propto}1/x^{\delta}$ $(G(x,Q^{2})=xg(x,Q^{2}))$, we
provide numerical results for the ratio $R(x{\rightarrow}
0,Q^{2}){\equiv}R_{\delta}(Q^{2})$ for values of the parameter
$\delta = 0$ or $\delta = 0.5$, the first value corresponds to the
soft Pomeron and the second value is corresponding to the hard
(Lipatov) Pomeron intercept. However, our analysis shows that the
predictions for $R^{c}$ with hard Pomeron intercept describe with
good accuracy the low $x$ predictions to NLO, and this analysis is
independent of the DGLAP evolution of the gluon PDF. In this
method, the charm structure function is determined without the
prior knowledge of the longitudinal charm
structure function.\\
Finally, our predictions show that the charm structure function
exponents have the same behaviour as the reduced charm cross
section exponents only for a particular range of $x$ values. The
structure of this article is as follows. In Sect.2 we present the
basic formalism of our approximation method with a brief review of
the calculational steps. The connection of  the ratio of the charm
structure functions with hard (Lipatov) Pomeron intercept is also
given. Later, we present the results for the charm structure
function with NLO corrections and show that our method reproduces
the HERA results for the charm structure function obtained by H1
Collaboration with the help of more cumbersome NLO estimations. In
Sect.3 we give the predictions for the charm structure function
exponents with respect to the reduced charm cross section
exponents at low $x$. These results are discussed in
Sect.4.\\

\section{2. The Hard (Lipatov) Pomeron Approach to the charm structure functions }

In the low- $x$ range, where the gluon contribution is dominant,
the charm quark contribution $F_{k}^{c}(x,Q^{2},m^{2}_{c})(k=2,L)$
to the proton structure function  is given by this form
\begin{eqnarray}
F_{k}^{c}(x,Q^{2},m^{2}_{c})&=&2e_{c}^{2}\frac{\alpha_{s}(\mu^{2})}{2\pi}\int_{1-\frac{1}{a}}^{1-x}dzC_{g,k}^{c}
(1-z,\zeta)\nonumber\\
&& {\times}G(\frac{x}{1-z},\mu^{2}),
\end{eqnarray}
where $a=1+4\zeta(\zeta{\equiv}\frac{m_{c}^{2}}{Q^{2}})$ and we
neglect the ${\gamma^{*}}q(\overline{q})$ fusion subprocesses.
 Here $G(x,\mu^{2})$ is the gluon distribution function and the mass
factorization scale $\mu$, which has been put equal to the
renormalization scale, is assumed to be either
$\mu^{2}=4m_{c}^{2}$ or $\mu^{2}=4m_{c}^{2}+Q^{2}$. In the above
expression $C^{c}_{g,k}$ is the charm coefficient function
expressed in terms of LO and NLO contributions as follows
\begin{eqnarray}
C_{k,g}(z,\zeta)&{\rightarrow}&C^{0}_{k,g}(z,\zeta)+a_{s}(\mu^{2})[C_{k,g}^{1}(z,\zeta)\\\nonumber
&&+\overline{C}_{k,g}^{1}(z,\zeta)\ln\frac{\mu^{2}}{m_{c}^{2}}],
\end{eqnarray}
where $a_{s}(\mu^{2})=\frac{\alpha_{s}(\mu^{2})}{4\pi}$ and in the
NLO analysis
\begin{eqnarray}
\alpha_{s}(\mu^{2})=\frac{4{\pi}}{\beta_{0}\ln(\mu^{2}/\Lambda^{2})}
-\frac{4\pi\beta_{1}}{\beta_{0}^{3}}\frac{\ln\ln(\mu^{2}/\Lambda^{2})}{\ln(\mu^{2}/\Lambda^{2})}
\end{eqnarray}
with $\beta_{0}=11-\frac{2}{3}n_{f},
\beta_{1}=102-\frac{38}{3}n_{f} $ ($n_{f}$ is the number of active
flavours).\\

In the LO analysis, the coefficient functions BGF can be found
[5-8], as
\begin{eqnarray}
C^{0}_{g,2}(z,\zeta)&=&\frac{1}{2}([z^{2}+(1-z)^{2}+4z\zeta(1-3z)-8{\zeta^{2}}z^{2}]\nonumber\\
&&{\times}\ln\frac{1+\beta}{1-\beta}+{\beta}[-1+8z(1-z)\nonumber\\
&&-4z{\zeta}(1-z)]),
\end{eqnarray}
and
\begin{eqnarray}
C^{0}_{g,L}(z,\zeta)=-4z^{2}{\zeta}\ln\frac{1+\beta}{1-\beta}+2{\beta}z(1-z),
\end{eqnarray}
where $\beta^{2}=1-\frac{4z\zeta}{1-z}$.\\
 At NLO,
$O(\alpha_{em}\alpha_{s}^{2})$, the contribution of the photon-
gluon component is usually presented in terms of the coefficient
functions $C_{k,g}^{1}, \overline{C}_{k,g}^{1}$.  The NLO
coefficient functions are only available as computer codes[9,10].
But in the high- energy regime ($\zeta<<1$) we can used the
compact form of these coefficients according to
the Refs.[11,12].\\

Applying the low- $x$ behaviour of the gluon distribution function
according to the hard (Lipatov) Pomeron [13-15]
\begin{equation}
G(x,\mu^{2}){\rightarrow}x^{-\delta}
\end{equation}
in Eq.5, integrating the gluon kernel over $z=\frac{x}{y}$ and
finally summing over the gluon distribution function yields
\begin{eqnarray}
F_{k}^{c}(x,Q^{2},m^{2}_{c})=2e_{c}^{2}\frac{\alpha_{s}(\mu^{2})}{2\pi}N_{k}(x,\mu^{2})\nonumber\\
{\times}G(x,\mu^{2}),
\end{eqnarray}
where
\begin{eqnarray}
N_{k}(x,\mu^{2})=\int_{1-\frac{1}{a}}^{1-x}C_{g,k}^{c}
(1-z,\zeta)(1-z)^{\delta}dz.
\end{eqnarray}
Therefore the ratio of the charm structure functions is given by
\begin{eqnarray}
R^{c}&=&\frac{F^{c}_L}{F^{c}_2}\nonumber\\
&&=\frac{N_{L}(x,\mu^{2})}{N_{2}(x,\mu^{2})},
\end{eqnarray}
In fact, this equation which is independent of  the gluon
distribution function, is very useful for practical
 applications.\\
 We insert this expression in Eq.4 to find
\begin{eqnarray}
\sigma^{c\overline{c}}_r=F^{^{c\overline{c}}}_2(x,Q^2,m^{2})(1-{\frac{y^2}{Y_+}}\frac{N_{L}(x,\mu^{2})}{N_{2}(x,\mu^{2})}),
\end{eqnarray}
or
\begin{eqnarray}
F^{^{c\overline{c}}}_2(x,Q^2,m^{2})=\sigma^{c\overline{c}}_r(1-{\frac{y^2}{Y_+}}\frac{N_{L}(x,\mu^{2})}{N_{2}(x,\mu^{2})})^{-1}.
\end{eqnarray}
This equation relates the charm structure function to the reduced
charm structure function via BGF kernels. We observe that the
right- hand side of Eq.15 is independent of the longitudinal charm
structure function and gluon distribution function, and this
formula can reproduce the HERA results for the charm structure
function from the reduced charm cross section.\\

\section{3. The charm structure function exponent behaviour }

A striking discovery at HERA [2] for the charm productions has
been the rapid rise of $\sigma^{c\overline{c}}_r$ with the energy
$W^2$ at each fixed $Q^{2}$ and small $x$ values. This rapid rise
is associated with the exchange of an object known as the hard
(Lipatov) Pomeron. In this powerful approach the charm structure
function has been found to have the same hard (Lipatov) Pomeron
behaviour [16-18]. This rise has been quantified by a study of the
observable
\begin{equation}
\lambda_{F^{{c\overline{c}}}_2}=\langle\frac{{\partial}{\ln}F^{{c\overline{c}}}_2}{{\partial}\ln1/x}\rangle
\end{equation}
where the brackets mean that this effective intercept
($\lambda_{F^{{c\overline{c}}}_2}$) is obtained from a fit of the
form
$F^{{c\overline{c}}}_2{\sim}x^{-\lambda_{F^{{c\overline{c}}}_2}}$
at fixed $Q^{2}$ and small  values of $x$. We analyse the H1
$\sigma^{c\overline{c}}_r$ data with a view to extracting such an
$x$ dependence. For this purpose, from Eq.4, we have the following
form
\begin{equation}
\langle\frac{{\partial}\ln\sigma^{{c\overline{c}}}_r}{{\partial}\ln1/x}\rangle=
\langle\frac{{\partial}{\ln}F^{{c\overline{c}}}_2}{{\partial}\ln1/x}\rangle+
\langle\frac{{\partial}\ln[1-\frac{y^{2}}{Y_{+}}R^{c}]}{{\partial}\ln1/x}\rangle.
\end{equation}
We would like remark that
\begin{equation}
\lambda_{\sigma^{{c\overline{c}}}_r}=\lambda_{F^{{c\overline{c}}}_2}+{\Delta}\lambda
\end{equation}
where
\begin{equation}
{\Delta}\lambda=\frac{R^{c}}{1-\frac{y^{2}}{Y_{+}}R^{c}}\frac{{\partial}}{{\partial}\ln1/x}\frac{y^{2}}{Y_{+}},
\end{equation}
or
\begin{eqnarray}
{\Delta}\lambda&=&-2{(2sx-Q^{2})R^{c}sQ^{4}x}[(2s^{2}x{^2}-2sxQ^{2}+Q^{4}-Q^{4}R^{c})\nonumber\\
&&{\times}(2s{^2}x{^2}-2sxQ^{2}+Q^{4})]^{-1} .
\end{eqnarray}
Here $s$ is the square of the total c.m. energy (which is constant
at HERA). As shown in Fig.1,  the $R^{c}$ behaviour is almost
independent of $x$ at all $Q^{2}$ values. ${\Delta}\lambda$  is
independent of the gluon distribution, depending only on the BGF
kernels. We shall analyse it for both small and large y data.\\
\section{4. Results and Conclusions }

We have analysed H1 data on charm production [3] and compared with
 DL
model [16-18] based on hard Pomeron exchange and also  with the
color dipole- model [19] and the GJR parametrisation [23]. Our
numerical predictions are presented as functions of $x$ for the
$Q^{2}=$12,20,35,60, 120 and 200 $GeV^{2}$. The $ep$ center of
mass energy is $\sqrt{s} = 319 GeV$, with a proton beam energy of
$E_{p} = 920 GeV$ and electron beam energy of $E_{e} = 27.6 GeV$
and also the average value $\Lambda$ in our
calculations  is corresponding to $224\hspace{0.1cm}MeV$.\\

In Figs.1 and 2 we show the predicted ratio of the charm structure
functions, $R^{c}$, as a function of $x$ and $Q^{2}$. As can be
seen from these figures $R^{c}$ is almost independent of $x$ for
$x<0.01$ in all $Q^{2}$ region. Also it is independent on the
choice of the gluon distribution function, where approaches based
on perturbative QCD and on $\textit{k}_{T}$ factorization give
similar predictions [20-22]. The effect of $R^{c}$ on the
corresponding differential charm cross- section should be
considered in extraction of $F_{2}^{c}$, so in Table 1 we give the
average of this ratio for various $Q^{2}$ values. We see
${\langle}R^{c}{\rangle}{\approx}0.1$ for a wide region of
$Q^{2}$. In principle, the parameter $\delta$ has two value
$\delta=0.08$ and $\delta= 0.44$. The first one is corresponds to
soft- pomeron exchange and the second one is corresponds to hard-
pomeron exchange. Our analysis shows that the predictions for
$R^{c}$ depend weakly on $\delta$, as it is less than $15\%$ in
the entire region of $Q^{2}$. Therefore, our approximation method
for $R^{c}$ with hard-pomeron behaviour describes with good
accuracy the low $x$ predictions for
$R^{c}$ when compared with results Ref.11 at NLO analysis (Fig.1 in Ref.11).\\

Now we use the analytic expression (15) for the extraction of the
charm structure function $F_{2}^{c}(x,Q^{2})$ from the $H1$
measurements of the reduced cross section using the NLO results
for $R^{c}$ derived in Section 2. Our results for the charm
structure function are presented in Table 2 and shown in Fig.3,
where they are compared with the values determined by the $H1$
analysis and with results obtained with the help of other standard
models (DL fit [16-18], color dipole model [19] and GJR
parametrization [23]). The agreement between our predictions with
the results obtained by H1 Collaboration  is remarkably good. Also
we observed that the theoretical uncertainty related to the
renormalization scales $\mu^{2}=4m_{c}^{2}$ and
$\mu^{2}=4m_{c}^{2}+Q^{2}$ is negligibly small. Our results for
extraction of the charm structure function from HERA measurements
of the reduced charm cross section are given for both the hard and
soft pomeron behaviour for $\delta$. One can see that the
predictions of our NLO analysis for the hard and soft pomeron
behaviour agree with the H1 data with an accuracy better than
$1\%$. This is because the contributions of the longitudinal charm
structure function to the reduced charm cross section is very
small. The numerical results show that a rough estimate of the
uncertainty in the charm structure function due to the difference
between the assumptions of a soft or a hard pomeron is less than $0.2\%$.\\

Finally, we analyse the behaviour of the exponents for the charm
structure function and the reduced charm cross section. To begin
with, we recognise that the behaviour of $\Delta\lambda$ (Eq.20)
is independent of the input gluon distribution function. The
behaviour of this expression as a function of $x$ (or $y$) is
shown in Fig.4 for several values of $Q^{2}$ ($Q^{2}$ and $s$  are
constant). We observe that the behaviour of $\Delta\lambda$
 at low $y$ (or high $x$) values is linear, ${\Delta}\lambda$ is very small so
that
$\lambda_{F^{{c\overline{c}}}_2}{\simeq}\lambda_{\sigma^{{c\overline{c}}}_r}$,
and this quantity can be determined from derivatives of the
$\sigma^{{c\overline{c}}}_r$ experimental data with respect of
$\ln1/x$ at $Q^{2}$ fixed. But in the large $y$ region (low $x$),
 ${\Delta}\lambda$ can no longer be neglected. The deviation of
this expression from zero shows the importance of non-linear
effects. Thus we observe that for large $x$ (low $y$ ) there are
not saturation effects. In this region the behaviour of the charm
structure function exponent
 and reduced charm cross section exponent is the same and the hard (Lipatov) pomeron picture gives a good fit to all data.
A depletion in the low $x$ (high $y$) is called shadowing whereas
an enhancement is called anti-shadowing [24].\\

The oscillating behaviour for ${\Delta}\lambda$ in Fig.4 can be
explained by new effects at low-$x$, such as the behaviour of the
gluon distribution evaluated with the nonlinear recombination. The
negative shadowing corrections to this behaviour can be explain by
the gluon recombination. This negative screening effect in the
recombination process originally occurs in the interferant cut-
diagrames of the recombination amplitudes. But the positive
anti-shadowing corrections comes from a general application of
momentum conservation. This anti- shadowing effect always coexists
with the shadowing effect in the QCD recombination processes.
Therefore we observe that significant non- linear effects to the
modified DGLAP equation for the charm structure function can be
apparent in the behaviour of the charm structure function
exponent. This shadowing and anti- shadowing effects in the
${\Delta}\lambda$ behaviour have different kinematic regions. The
net effects for ${\Delta}\lambda$ behaviour depends not only on
the size of gluon distribution at this value of $x$ in the
modified DGLAP equation for the charm structure function, but also
on the shape of the gluon distribution when the Bjorken variable
goes from $x$ to $x/2$. In consequence, the shadowing effects in
the exponents behaviour will be obviously weakened by the
antishadowing effects as the gluon distribution has a  steeper
behavior at low-$x$ values. This transition for ${\Delta}\lambda$
is shown in Fig.4. The main prediction is a transition from the
linear
 regime described for the exponents to a non-linear regime where parton recombination
  becomes important in the parton cascade. In view of these results for the
 exponents, we may infer some evidence for non- linear effects at
 HERA and RHIC.\\

 In summary, we have used the hard (Lipatov) pomeron for the low
 $x$ gluon distribution to predict the charm structure functions ($F_{2}^{c}$ and $F_{L}^{c}$).
We derived a compact formula for the
 ratio
 $R^{c}=\frac{F^{^{c\overline{c}}}_L}{F^{^{c\overline{c}}}_2}$ that is valid through NLO at small
  values of Bjorken$^{,}$s $x$ variable, as it is independent of the input gluon
 distribution function. We have checked that this formula gives a good
description of the
 charm structure function from the reduced charm cross section data without the prior
knowledge of the longitudinal charm structure function. Careful
investigation of our results show a good agreement with the
previously published charm structure functions and other models.
Then we have used it to predict the behaviour of
  the charm structure function and reduced charm cross
 section exponents in the Perturbative $Q^{2}$ region at small
 $x$. Our predictions indicate that the behaviour of ${\Delta}\lambda$ can be explained by  non linear effects at
very low $x$. These numerical results show that the correction of
 the modified DGLAP equation due to gluon recombination
incorporates both shadowing and antishadowing effects, as the
influence of the antishadowing effect to the pre-asymptotic form
of the gluon distribution is non-negligible.\\
\acknowledgments G.R.Boroun thanks Prof.A.Cooper-Sarkar for
interesting and useful discussions.

\begin{table}
\caption{The predictions for the ratio
$R^{c}=\frac{F_{L}^{c}}{F^{c}_{2}}$ as a function of $Q^{2}$
corresponding to the cases of hard and soft pomeron.}
\label{tab.1}
\begin{center}
\begin{tabular}{|l|c|c|c|} \hline\noalign{\smallskip} $Q^{2}(GeV^{2})$ & $<R^{c}>(\delta\simeq 0.5)$ & $<R^{c}>(\delta\simeq 0)$ \\
\hline\noalign{\smallskip}
12 & 0.078 & 0.066  \\
20 & 0.092  & 0.078  \\
35 & 0.103   & 0.089 \\
60 & 0.109    & 0.094\\
120 & 0.111   & 0.097 \\
200 & 0.109   & 0.096 \\
300 & 0.106 & 0.094\\
600 & 0.101 & 0.090\\
1000 & 0.096 & 0.086\\
10000 & 0.078 & 0.071\\
100000 & 0.065 & 0.061\\
\hline\noalign{\smallskip}
 \end{tabular}
\end{center}
\end{table}
\begin{figure}
\includegraphics[width=0.52\textwidth]{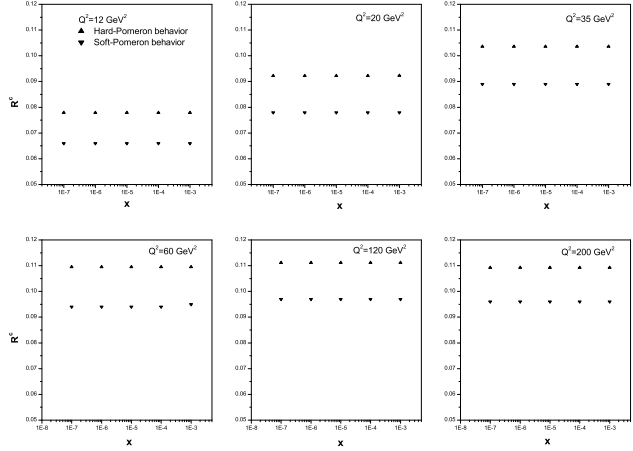}
\caption{The ratio
$R^{c}=\frac{F^{^{c\overline{c}}}_L}{F^{^{c\overline{c}}}_2}$ as a
function of $x$ for different values of $Q^{2}$ at $\delta\simeq0$
and $\delta\simeq 0.5$.}
\end{figure}
\begin{figure}
\includegraphics[width=0.51\textwidth]{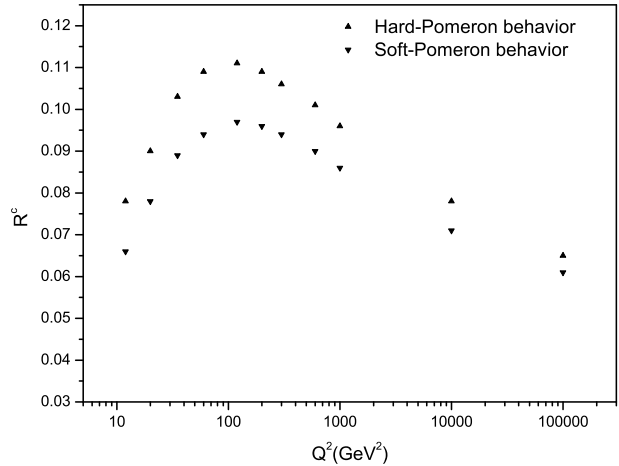}
\caption{The ratio
$R^{c}=\frac{F^{^{c\overline{c}}}_L}{F^{^{c\overline{c}}}_2}$ as a
function of $Q^{2}$ values at $\delta\simeq0$ and $\delta\simeq
0.5$.}
\end{figure}
\begin{figure}
\includegraphics[width=0.5\textwidth]{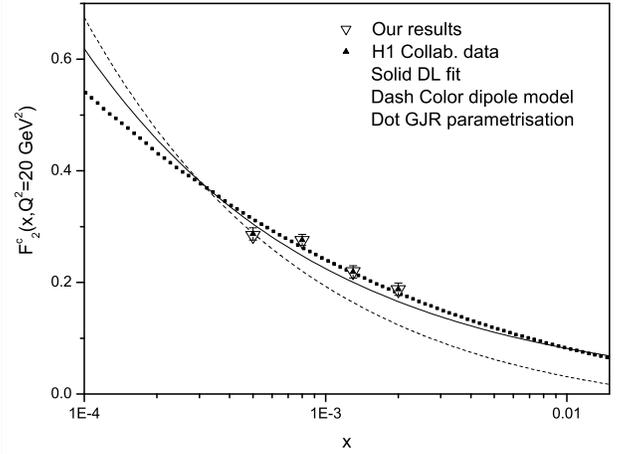}
\caption{The structure function $F_{2}^{c}(x,Q^{2})$ as a function
of $x$ at $Q^{2}=20 GeV^{2}$ compared with HERA data that
accompanied with total errors (F.D. Aaron et
al.,Eur.Phys.J.C\textbf{65},89(2010)) [3], DL fit [16-18], color
dipole model[19] and GJR parametrisation [23].}
\end{figure}
\begin{figure}
\includegraphics[width=0.55\textwidth]{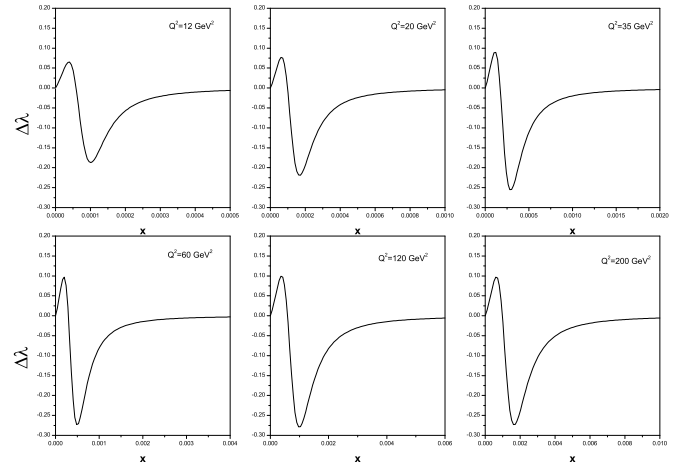}
\caption{The nonlinear behaviour of the exponents (
${\Delta}\lambda=\lambda_{F^{{c\overline{c}}}_2}-\lambda_{\sigma^{{c\overline{c}}}_r}$)
as a function of $x$ at $Q^{2}$ values.}
\end{figure}

\newpage
\begin{table}
 \caption{The values of $F_{2}^{c}(x,Q^{2})$
corresponding to the hard-pomeron behaviour extracted from the
$\widetilde{\sigma}^{c}(x,Q^{2})$ at low and high $Q^{2}$ for
various values of $x$  are compared with H1 results [3] that
accompanied with total errors ($\Delta \%$). } \label{tab.1}
\begin{center}
\begin{tabular}{|l||c|c||c|c||c|c||c|c|} \hline\noalign{\smallskip} $Q^{2}(GeV^{2})$ & $ x$ &
$ y$ & $ \widetilde{\sigma}^{c\overline{c}}$ & $
\Delta_{\widetilde{\sigma}^{c\overline{c}}}(\%)$ &
$F_{2}^{c\overline{c}}$(Ref.3) &$
\Delta_{F_{2}^{c\overline{c}}}(\%)$ & $F_{2}^{c\overline{c}}$($\delta \simeq 0.5$)& $F_{2}^{c\overline{c}}$($\delta \simeq 0$) \\
\hline\noalign{\smallskip}
12 & 0.00130 & 0.091 & 0.150 & 18.7& 0.150 & 1.0 &0.1500 & 0.1500   \\
12 & 0.00080 & 0.148 & 0.177 & 15.9& 0.177 & 1.1 &0.1772 &  0.1771 \\
12 & 0.00050 & 0.236 & 0.240 & 11.2& 0.242 & 1.0 &0.2407  & 0.2405 \\
12 & 0.00032 & 0.369 & 0.273 & 13.8& 0.277 & 1.1 &0.2751  & 0.2747 \\
20 & 0.00200 & 0.098 & 0.187 & 12.7 & 0.188 & 1.1 &0.1871 & 0.1871\\
20 & 0.00130 & 0.151 & 0.219 & 11.9 & 0.219 & 1.1 &0.2193 &  0.2192\\
20 & 0.00080 & 0.246 & 0.274 & 10.2 & 0.276 & 1.0 &0.2750 & 0.2748\\
20 & 0.00050 & 0.394 & 0.281 & 13.8 & 0.287 & 1.1 &0.2840 & 0.2834\\
35 & 0.00320 & 0.108 & 0.200 & 12.7& 0.200 & 1.1&0.2001   & 0.2001\\
35 & 0.00200 & 0.172 & 0.220 & 11.8& 0.220 & 1.0&0.2204 & 0.2203\\
35 & 0.00130 & 0.265 & 0.295 & 9.70& 0.297 & 1.0&0.2964 & 0.2962\\
35 & 0.00080 & 0.431 & 0.349 & 12.7& 0.360 & 1.1&0.3541 & 0.3533\\
60 & 0.00500 & 0.118 & 0.198 & 10.8& 0.199 & 1.1&0.1982 & 0.1981\\
60 & 0.00320 & 0.185 & 0.263 & 8.40 & 0.264 & 1.0&0.2636 & 0.2635\\
60 & 0.00200 & 0.295 & 0.335 & 8.80 & 0.339 & 1.0&0.3372 & 0.3368\\
60 & 0.00130 & 0.454 & 0.296 & 15.1& 0.307 & 1.0&0.3012 & 0.3004\\
120& 0.01300 & 0.091 & 0.133 & 14.1& 0.133 & 1.2&0.1331 & 0.1330\\
120& 0.00500 & 0.236 & 0.218 & 11.1& 0.220 & 1.1&0.2190 & 0.2187\\
120& 0.00200 & 0.591 & 0.351 & 12.8& 0.375 & 2.9&0.3630 & 0.3612\\
200& 0.01300 & 0.151 & 0.161 & 11.9& 0.160 & 2.7&0.1602 & 0.1612\\
200& 0.00500 & 0.394 & 0.237 & 13.5& 0.243 & 2.9&0.2400 & 0.2396\\
300& 0.02000 & 0.148 & 0.117 & 18.5& 0.117 & 2.9&0.1171 & 0.1171\\
300& 0.00800 & 0.369 & 0.273 & 12.7& 0.278 & 2.9&0.2760 & 0.2755\\
\hline\noalign{\smallskip}
 \end{tabular}
\end{center}
\end{table}

%

\end{document}